\documentclass[journal]{IEEEtran}

\usepackage[utf8]{inputenc}
\usepackage{cite}
\usepackage{float}
\usepackage{multicol,multirow}
\usepackage{makecell,booktabs}
\usepackage{hyperref}
\usepackage{graphicx}
\usepackage{amsfonts}
\usepackage{amssymb}
\usepackage{amsmath}
\usepackage{array}
\newcolumntype{P}[1]{>{\centering\arraybackslash}p{#1}}
\usepackage{color}
\usepackage{epstopdf}
\usepackage{subcaption}
\usepackage{bbm}
\usepackage{bm}
\usepackage{comment}
\usepackage{forest}
\usepackage{nomencl}
\usepackage{dblfloatfix}

\usepackage{etoolbox}

\usepackage{algorithm}
\usepackage[noend]{algpseudocode}
\makeatletter
\def\BState{\State\hskip-\ALG@thistlm}
\makeatother
\algdef{SE}[DOWHILE]{Do}{doWhile}{\algorithmicdo}[1]{\algorithmicwhile\ #1}
\renewenvironment{IEEEbiography}[1]
  {\IEEEbiographynophoto{#1}}
  {\endIEEEbiographynophoto}

\begin{document}
    \title{A P2P-dominant Distribution System Architecture}
    \author{Jip~Kim,~\IEEEmembership{Student Member,~IEEE,}        
        and~Yury~Dvorkin,~\IEEEmembership{Member,~IEEE}  
    \thanks{The authors are with the Department of Electrical and Computer Engineering, Tandon School of Engineering, New York University, NY 11201 USA. This work was supported by U.S. NSF under Grant ECCS-1847285.}
    }  

    \maketitle

    \begin{abstract}        
        Peer-to-peer interactions between small-scale energy resources exploit distribution network infrastructure as an electricity carrier, but remain financially unaccountable to electric power utilities. 
        This status-quo raises multiple challenges. First, peer-to-peer energy trading reduces the portion of electricity supplied to end-customers by utilities and their revenue streams. Second, utilities must ensure that peer-to-peer transactions comply with distribution network limits. This paper proposes a peer-to-peer energy trading architecture, in two configurations, that couples peer-to-peer interactions and  distribution network operations. The first configuration assumes that these interactions are settled by the utility in a centralized manner, while the second one is peer-centric and does not involve the  utility. Both configurations use distribution locational marginal prices to compute network usage charges that peers must pay  to the utility for using the distribution network.
    \end{abstract}
    

    \section*{Nomenclature}
    
    \subsection{Sets and Indices}
    \addcontentsline{toc}{subsection}{Sets and Indices}
    \begin{IEEEdescription}[\IEEEusemathlabelsep\IEEEsetlabelwidth{$\Lambda_\omega\!=\!\{\rho_\omega^{\mathrm b},\rho_\omega^{\mathrm s}\}\!\!\!$}]
    \item[$b\in {\cal{B}}$]{Set of buses}
    \item[$l\in {\cal{L}}$]{Set of distribution lines}
    \item[$n\in {\cal{N}}$]{Sets of peers where ${\cal N}^{\mathrm{b/s}}$ denote sets of buying/selling peers, ${\cal N}^{\mathrm b}\!\cup\! {\cal N}^{\mathrm s}\!=\! {\cal N}$}
    \item[$\omega\in {\Omega}$]{Set of peer trades}
    \item[$\omega \in {\Omega}_n$]{Sets of trades of peer $n$,  $\bigcup_{n\in {\cal N}}\Omega_n = \Omega$}
    \item[$\omega\in {\Omega^*}$]{Set of the matched peer trades, $\Omega^* \subseteq \Omega$}
    \item[$\Lambda_\omega\!=\!\{\rho_\omega^{\mathrm b},\rho_\omega^{\mathrm s}\}$]{Set of buying/selling prices of trade $\omega$}
    \item[$b(\omega)/s(\omega)$]{Indices of buying/selling peers in trade $\omega$}
    \item[$o(l)/r(l)$]{Indices of originating/receiving-end nodes of distribution line $l$}
    \end{IEEEdescription}

    \subsection{Parameters}
    \addcontentsline{toc}{subsection}{Parameters}
    \begin{IEEEdescription}[\IEEEusemathlabelsep\IEEEsetlabelwidth{$C_{s(\omega)}($}]
    \item[$B_b$]{Susceptance of bus $b$ [$\mathrm{p.u.}$]}
    \item[$C_{n}(\cdot)$]{Cost function of selling peer $n$ [$\$/\mathrm{MWh}$]}
    \item[$C^{\mathrm u}_b$]{Cost of utility generator at bus $b$ [$\$/\mathrm{MWh}$]}
    \item[$C^{\mathrm w}$]{Wholesale market electricity price [$\$/\mathrm{MWh}$]}
    \item[$D^{\mathrm{p}}_{b}/D^{\mathrm{q}}_{b}$]{Active/reactive power demand [$\mathrm{MW/MVAr}$]}
    \item[$\underline{D}_n^{\mathrm{p}}/\overline{D}_n^{\mathrm{p}}$]{Minimum/maximum power bought by peer $n$ [$\mathrm{MW}$]\!\!\!\!}
    \item[$G_b$]{Conductance of bus $b$ [$\mathrm{p.u.}$]}     
    \item[$\underline{G}_n^{\mathrm{p}}/\overline{G}_n^{\mathrm{p}}$]{Minimum/maximum power sold by peer $n$ [$\mathrm{MW}$]}
    \item[$P$]{Standard trade size [$\mathrm{MW}$]}    
    \item[$\underline{P}_b^{\mathrm{g}}/\overline{P}_b^{\mathrm{g}}$]{Minimum/maximum real power output limit of utility generator at bus $b$ [$\mathrm{MW}$]}
    \item[$\underline{Q}_b^{\mathrm{g}}/\overline{Q}_b^{\mathrm{g}}$]{Minimum/maximum reactive power output limit of utility generator at bus $b$ [$\mathrm{MVAr}$]}
    \item[$R_l$]{Resistance of distribution line $l$ [$\mathrm{p.u.}$]}
    \item[$S_l$]{Apparent flow limit of distribution line $l$ [$\mathrm{MVA}$]}
    \item[$T_b$]{Electricity tariff at bus $b$ [$\mathrm{\$/MWh}$]}
    \item[$U_n(\cdot)$]{Utility function of buying peer $n$ [$\$/\mathrm{MWh}$]}
    \item[$\underline{V}_b/\overline{V}_b$]{Minimum/maximum limit on  the squared voltage magnitude at bus $b$ [$\mathrm{p.u.}$]}
    \item[$X_l$]{Reactance of distribution line $l$ [$\mathrm{p.u.}$]}
    \item[$\Gamma_{b}$]{Penetration level of the peer trading at bus $b$}    
    \item[$\Pi^{\mathrm u}$]{Revenue of the power utility [$\$$]}
    \item[$\Upsilon_n$]{Value of electricity surplus for peer $n$ [$\$/\mathrm{MW}$]}
    \item[$\Delta \rho$]{Value of adjustment in price [$\$/\mathrm{MWh}$]}
    \end{IEEEdescription}

    \subsection{Variables}    
    \addcontentsline{toc}{subsection}{Variables}
    \begin{IEEEdescription}[\IEEEusemathlabelsep\IEEEsetlabelwidth{$\eta^{\mathrm{+}}_l/\eta^{\mathrm{-}}_l\!$}]
    \item[$a_{l}$]{Squared current flow of distribution line $l$ [$\mathrm{p.u.}$]}
    \item[$c^{\mathrm n}_{\omega}$]{Network usage charge for trade $\omega$ [$\$/\mathrm{MWh}$]}    
    \item[$d^{\mathrm{p}}_{n}$]{Power bought by peer $n$ [$\mathrm{MW}$]}
    \item[$f^{\mathrm{p}}_{l}/f^{\mathrm{q}}_{l}$]{Active/reactive power flow of distribution line $l$ [$\mathrm{MW/MVAr}$]}    
    \item[$g^{\mathrm{p}}_{n}$]{Power sold by peer $n$ [$\mathrm{MW}$]}
    \item[$p^{\mathrm{g}}_{b}/q^{\mathrm{g}}_{b}$]{Active/reactive  power output of utility generator at bus $b$ [$\mathrm{MW/MVAr}$]}
    \item[$p_{\omega}$]{Power transfer from $s(\omega)$ to $b(\omega)$ in trade $\omega$ [$\mathrm{MW}$]}
    \item[$v_{b}$]{Squared nodal voltage magnitude of bus $b$ [$\mathrm{p.u.}$]} 
    \item[$\eta^{\mathrm{+}}_l/\eta^{\mathrm{-}}_l$]{Dual variable of forward/backward flow limit constraints on distribution line $l$}
    \item[$\lambda_b$]{Dual variable of the active power balance constraint at bus $b$}
    \item[$\mu_b$]{Dual variable of the reactive power balance constraint at bus $b$}
    \item[$\rho_\omega^{\mathrm b}/\rho_\omega^{\mathrm s}$]{Buying/selling price of trade $\omega$ [$\$/\mathrm{MWh}$]}
    \end{IEEEdescription}
    
    \section{Introduction} \label{sec:intro}    
        Owing to  recent advances in smart grid technologies, the U.S. power grid is undergoing   nation-wide modernization. One of the most important objectives of this modernization is to achieve a high degree of supply autonomy  of electricity consumers from their local electric power utility and the freedom to choose their electricity suppliers.
        In practice, the supply autonomy and the freedom to choose are enabled by rolling out customer-end distributed energy resources (DERs), which include, but are not limited to, photovoltaic panels, battery energy storage, and demand-side management. If these DERs are appropriately sized and operated, electricity consumers are shown to significantly reduce, if not completely eliminate, their dependency on the electricity supply from the utility. 
        
        While the roll-out of DERs offers significant reliability and economic benefits to both the utility and consumers, it reduces revenue streams of the utility and undermines their financial viability. Furthermore, accommodating large-scale DER deployment also imposes technical challenges on the distribution network operations since the current electric power distribution  infrastructure was not designed to deal with bidirectional power flows, increased voltage fluctuations and volatile nodal power injections induced by DERs. As a response to these challenges, utilities in many U.S. regions have already started increasing electricity tariffs, thus further incentivizing remaining consumers to adopt DERs and exacerbating their impact on the distribution system \cite{seattle}. This self-fueling process -- colloquially known as the utility's death spiral -- calls for urgent changes to the current electric power distribution practice. Accordingly, 94\% of the senior power and the utility executives surveyed by PricewaterhouseCoopers predict `\textit{complete transformation [...] to the power utility business model}' by 2030, \cite{pwc_2013}. 
        
        These techno-economic challenges observed by utilities motivate to re-think and re-engineer interactions between stand-alone DERs and utilities to continue harvesting their benefits without compromising supply reliability. Among possible alternatives, the peer-to-peer (P2P) architecture is regarded as a viable coordination mechanism that can efficiently operate heterogeneous DERs, \cite{nature_prosumer_era}, while respecting physical limits on the distribution network. The P2P architecture assumes a less centralized, more autonomous and flexible electricity delivery, in which small-scale (e.g. residential and commercial) producers and consumers  can   transact electricity and other services  as an alternative to centralized electricity supply from utilities or third-party aggregators.
        
        Unlike the current  distribution  practice that mainly pursues the economies of scale and scope benefits \cite{BURGER2017395}, the value proposition of the P2P architecture  stems from the sharing economy \cite{Demary2015Competition}. The sharing economy monetizes under-utilized or otherwise suboptimally used resources due to the failure of utilities and aggregators to effectively communicate with and aggregate DERs, \cite{jowskow, BURGER2017395}. 
        However, the current regulatory environment does not incentivize power utilities to accommodate P2P, thus hindering their value to the system \cite{mit_utility_of_the_future2, BURGER2017395}.
        
        The literature on P2P interactions in distribution systems is thin and still emerging. Morstyn \textit{et al.} \cite{morstyn2019bilateral,morstyn2018using,morstyn2018multi} leverage the concept of full substitutability \cite{full_substit}, i.e. an equilibrium condition for peers in a hierarchical supply chain, to develop a bilateral contract network for P2P energy trading. This contract network allows for forward (e.g. day-ahead) and real-time energy trading that produce a time-static, network-unconstrained stable equilibrium that peers have no incentive to deviate from. Park \textit{et al.} \cite{park2016contribution} derived a closed form of network-unconstrained Nash equilibrium among peers in microgrids and Tushar \textit{et al.} \cite{8398582} review game- and auction-theoretic approaches to represent the P2P interactions under different implementation scenarios.         
        Relative to \cite{full_substit,park2016contribution}, the authors of \cite{8398582} generalize the definition of the P2P equilibrium  in the network-constrained context. While \cite{8398582} emphasizes the importance of accounting for network constraints in P2P transactions, it does not describe how it can be done. 
        To account for possible network limitations, Ahn \textit{et al.} \cite{ahn2018distributed} restrict P2P interactions to neighboring nodes and exploit the Lagrangian duality to compute the electricity  and ``energy flow'' prices. The latter price is used to charge peers for using the network infrastructure operated by utilities, but does not capture the effect of P2P interactions in other (more remote) distribution grid locations.  
        This limitation is partly addressed by Baroche \textit{et al.} \cite{baroche2018exogenous}, where the authors consider P2P energy transactions across a given network and account for basic power flow constraints via the DC power flow approximation. The model in \cite{baroche2018exogenous} can also enforce a fixed, exogenous charge on peers for using the utility's network infrastructure. However, the use of the DC approximation does not reflect the distribution network physics (e.g. losses, voltage regulation, reactive power support) and, thus, the actual cost incurred by the utility. Furthermore, exogenously set charges in \cite{baroche2018exogenous} do not capture spatio-temporal dynamics of the distribution system and sensitivities of peers.
        M{\"u}nsing \textit{et al.} \cite{munsing2017blockchains} present a blockchain-enabled decentralized P2P market-clearing algorithm using a convexified AC optimal power flow (OPF) and exploit the distribution locational marginal prices (DLMPs) to settle transactions among peers. Building on \cite{munsing2017blockchains}, Wang \textit{et al.} \cite{wang2018blockchain,wang2019energy} leverage blockchains to design a ``Crowdsourced energy system" with a P2P energy trading for day-ahead and real-time operations. 
        Although \cite{munsing2017blockchains,wang2018blockchain,wang2019energy} incentivize P2P interactions, they do not compensate the power utilities for peers' usage of  the distribution network.
        
        The P2P matching mechanisms, i.e.  methods to connect peering producers and consumers,  can be classified as either  system-centric{\cite{8398582, ahn2018distributed, baroche2018exogenous,munsing2017blockchains,wang2018blockchain,wang2019energy}} or peer-centric{\cite{morstyn2018using,morstyn2018multi,morstyn2019bilateral,park2016contribution}}. The system-centric matching resembles pool-structured wholesale electricity markets with the single supervisory entity that collects and matches the bids and offers submitted by market participants in a centralized manner. On the other hand, the peer-centric approach is  decentralized, which offers more flexibility for accommodating their preferences, \cite{full_substit}, and allows for distributed decision-making protocols that preserve privacy of peers, \cite{sorin2019consensus,guerrero2018decentralized}.  Regardless of the matching mechanism chosen, the  large-scale implementation of P2P interactions is expected to affect the ability of the utility to operate the distribution network efficiently and reliably.
        
        This paper aims to {design} a new electric power distribution  architecture that will allow for a large volume of P2P interactions among small-scale DERs and shift electric power  utilities from the current volumetric business model, when the revenue is proportional to the amount of electricity sold to customers, to a service-based business model, where the revenue is collected from providing services to support electricity trading by other parties.         
        { To this end, we conceptualize the P2P platform{ \footnote{ Our definition of the P2P platform is technology agnostic, e.g. it can be enabled by either blockchains \cite{park2016contribution,munsing2017blockchains,wang2018blockchain,wang2019energy} or other decentralized technologies.}}, 
        i.e.  a marketplace for direct energy transactions among peers, that can internalize both the system- and peer-centric matching approaches.}  To effectively  accommodate P2P in the distribution system, the P2P platform is then integrated with a distribution AC OPF. {This integration aims to capture the sharing economy benefits, without compromising supply reliability, and ensures that P2P interactions are accounted for in OPF-based energy management tools.         
        Unlike \cite{munsing2017blockchains,wang2018blockchain,wang2019energy}, this paper uses the OPF framework to derive and compute network usage charges to be paid by peers for using the distribution network based on the DLMPs, \cite{Papavasiliou:2017ek}.} The use of the DLMPs makes it possible to represent spatio-temporal dimensions of operating conditions in the distribution network and consider them while computing the network usage charges. In turn, the network usage charges  are then used to encourage those P2P transactions that improve the overall distribution system performance and generate an  additional revenue stream intended to offset the drop in the utility's revenue caused by the  roll-out of customer-end DERs.

\section{Distribution System with the P2P Platform}\label{sec:utility_model}

Regardless of the peer matching mechanism chosen, it is important to consider generic interfaces that relates the P2P platform and the rest of the distribution system. Fig.~\ref{fig:platform} illustrates these interfaces for different volumes of the electricity supplied by the P2P platform as compared to the current distribution system architecture.

    \subsection{Current distribution system architecture}\label{subsec:status-quo}
    The current distribution system architecture is shown in Fig.~\ref{fig:platform}(a), where the sole utility operates the distribution network, supplies electricity to customers and collects the electricity payment. 
     Under this practice, the electricity prices for small-scale consumers are based on flat or time-of-use volumetric electricity rates, which are typically regulated and set to recover both the operating and capital costs incurred by the utility.  The operating cost includes the cost of electricity supply, maintenance, network losses and  control, while the capital cost includes the cost of expansion and upgrade projects. In this case, the revenue of the utility is:
    \begin{align}
        &\Pi^{\mathrm{u}}=\sum_{b\in{\cal B}} T_b D^{\mathrm{p}}_b, \label{DSOrev_cen} 
    \end{align}

    \noindent and is proportional to  rate $T_b$ and  active power demand $D^{\mathrm{p}}_b$, thus assuming all operating and capital costs are uniformly allocated among customers based on their electricity consumption. However, this approach fails to adequately capture the costs incurred by the customers that deployed their own DERs, which reduce or eliminate their electricity consumption provided by the utility and thus the utility revenue.  Since rate $T_b$ lumps together different operating and capital costs, it is impossible to accurately itemize the effect of DERs on the operating cost.   Therefore, the current practice impedes further proliferation of DERs because it does not provide sufficient compensation for network services provided by the utility. 

    \textcolor{black}{While the time-of-use rates recognize temporal (e.g. intra-day) fluctuations of electricity demand, their temporal granularity is fairly coarse and usually is limited to two intra-day internvals (e.g., peak and off-peak rate). Furthermore, their spatial granularity does not recognize network peculiarities of electric power distribution systems and is typically set on municipality boundaries.  As an attempt to increase spatio-temporal granularity of tariffs, Caramanis \textit{et al.} \cite{7429676} proposed to introduce DLMPs that would internalize these network peculiarities and dynamically changing demand conditions in the price formation process (similarly to wholesale locational marginal prices), thus improving pricing fidelity. The DLMPs have been shown to accurately reflect the physics of AC power flows in distribution systems, \cite{Papavasiliou:2017ek}, and can be extended to accommodate uncertain nodal injections, \cite{mieth2019distribution}. However, in practice, DLMPs have not been implemented yet, in part due to lacking advanced metering infrastructure and socio-economic implications that granular electricity prices may cause, \cite{8122642}. }
    
    \subsection{Distribution system architecture with the P2P platform}\label{subsec:part_decen}        
    
    \begin{figure}[!t]
        \centering
        \includegraphics[width=\columnwidth]{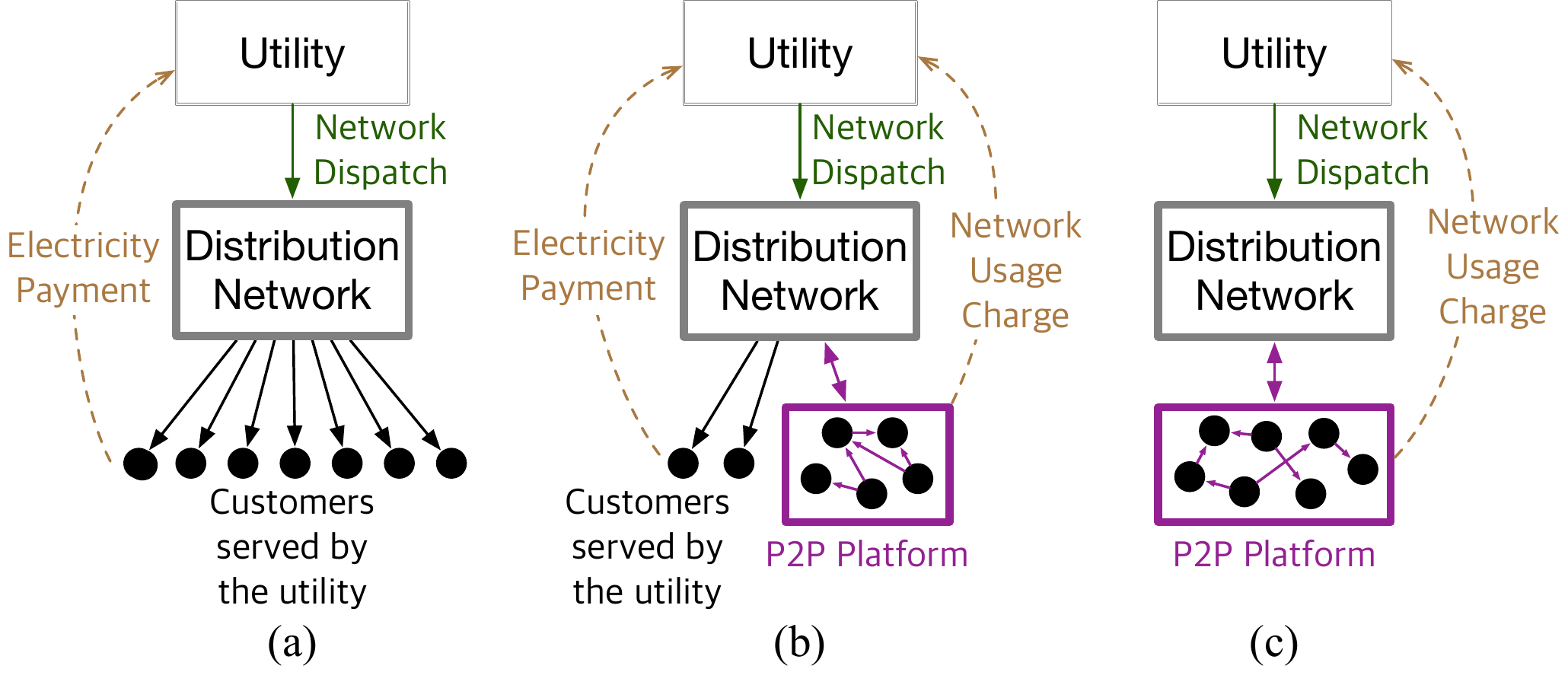}
        \caption{\small Comparison of the distribution system architecture with and without the proposed P2P platform: (a) Current architecture, (b) Mixed architecture, (c) P2P architecture. { Solid and dashed arrows represent energy and financial flows respectively.}}
        \label{fig:platform}
    \end{figure}

    Based on the open access assumption, we envision two possible distribution architectures that can replace the current architecture in Fig.~\ref{fig:platform}(a). The mixed architecture illustrated in  Fig.~\ref{fig:platform}(b) preserves the current option of receiving electricity supply from the utility for some consumers and also enables the P2P interactions among stand-alone DERs and consumers. Under the mixed architecture, the P2P platform matches producing and consuming peers and sets the electricity price for them. Since the P2P transactions will rely on the utility to operate the network, the peers are also additionally charged by the utility for using the network infrastructure. Therefore, the utility  revenue is given as follows:
    \begin{align}
        &\Pi^{\mathrm{u}}\!=\sum_{b\in{\cal B}} T_b D^{\mathrm{p}}_b (1\!-\!\Gamma_b)  \!+\!\! NUC( c_{\omega}^{\mathrm{n}}, p_\omega), \label{DSOrev_partial-decen}
    \end{align}

    \noindent where the first term represents the electricity payment from the customers that receive their electricity supply directly from the utility and {$NUC(\cdot)$ is the total  network usage charges collected by the utility from the peers participating in the P2P interactions.} 
    In Eq.~\eqref{DSOrev_partial-decen}, parameter $\Gamma_b\in [0,1]$ defines the ratio between the total demand of a peer located at bus $b$ served by the utility and by the P2P platform. Thus,  the total demand procured by the P2P platform is $\sum_{\omega\in\Omega^*}p_\omega = \sum_{b \in \mathcal{B}} D^{\mathrm p}_b \Gamma_b$.

    The P2P architecture in Fig.~\ref{fig:platform}(c) represents a particular case of the mixed architecture in Fig.~\ref{fig:platform}(b), where the utility does not supply electricity and only supports network operations and the P2P platform satisfies the demand of all customers. In this case the utility revenue can be obtained by setting $\Gamma_b=1$:
    \begin{align}        
        &\Pi^{\mathrm{u}}\!= NUC( c_{\omega}^{\mathrm{n}}, p_\omega). \label{DSOrev_full-decen}
    \end{align}         
    \noindent The key component of both the mixed and P2P architectures described above is network usage charge $NUC( c_{\omega}^{\mathrm{n}}, p_\omega)$ that needs to be designed to recover the cost incurred by the utility while operating the distribution network and factored in the price formation process within the P2P platform. Similarly to the DLMPs, the network usage charge in  Eq.~\eqref{DSOrev_full-decen} is computed based on operating conditions and does not aim to recover capital (long-term) costs.

\section{P2P Trading with Network Usage Charge}\label{sec:platform}

Implementing the P2P architecture, as shown in Fig.~\ref{fig:platform}(b)-(c), requires routines to enable the peer matching process and to couple the P2P interactions with distribution network operations. Section~\ref{subsec:peermatching} describes two distinct peer-matching routines, while Sections~\ref{subsec:NUC} and \ref{subsec:NUC_integration} introduce network usage charges to integrate these routines in one decision-making process with distribution network operations. 
    
    \begin{figure}[t]
        \centering
        \includegraphics[width=0.95\columnwidth]{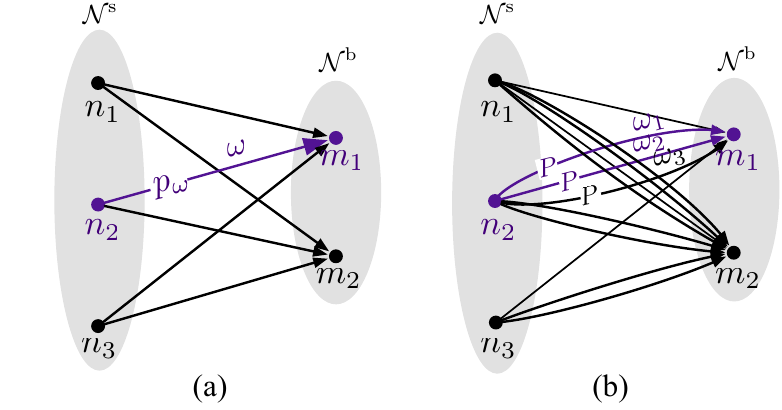}
        \caption{\small A schematic representation of the peer matching process as (a) a simple bipartite graph for the system-centric configuration and (b) a bipartite graph with parallel edges for the peer-centric configuration. The edge represents trade $\omega$ of power $p_{\omega}$ between seller $s(\omega)$ and consumer $b(\omega)$.}
        \label{fig:bipartite}                
    \end{figure}

    \subsection{Peer matching process} \label{subsec:peermatching}
    We develop the peer-matching routines for two possible configurations that may arise in the future. The first configuration, referred to as system-centric, assumes that the utility will be responsible for matching the peers in a welfare-maximizing manner. By contrast, the peer-centric routine is carried out autonomously from the utility and is driven by preferences and choices of the peers. 
    For the sake of modeling convenience, both the system- and peer-centric configurations described below assume that each peer can be either a producer, or a consumer, or an idler at a given time instance. This role assignment can also change at different time instances. For example, if a peer is equipped with an energy storage unit, it can act as a producer, when it discharges power to the distribution system, and as a consumer, when it charges power from the distribution system.
    Also, there are no exogenous restrictions on the set of possible matches between the consuming and producing peers.
    Under these two assumptions, the peer matching process in both configurations can be represented as a bipartite graph in Fig.~\ref{fig:bipartite}. 
    Each P2P energy trade $\omega$ is represented as an edge with sending and receiving nodes $n\in{\cal N}^{\mathrm s}$ and $m\in{\cal N}^{\mathrm b}$, respectively. We also denote node $n$ of edge $\omega$ as seller $s(\omega)\!=\!n$ and node $m$ as buyer $b(\omega)\!=\!m$ for every  trade $\omega$ with the traded power denoted as $p_\omega$.    
    The system-centric configuration is represented as a simple (trivial) bipartite graph in Fig.~\ref{fig:bipartite}(a), i.e. there is only one edge (trade) connecting a buying and a selling peer. Therefore, the number of potential matches to be considered is $\vert \Omega\vert\!=\! \text{card}({\cal N}^{\mathrm s})\times\text{card}({\cal N}^{\mathrm b})$, and the objective of the matching process is to determine power transfer $p_\omega$ between peers given in trade set $\Omega$. On the other hand in the peer-centric configuration that allows peers to negotiate trade prices, all trades have  standard size  $p_\omega \!=\! P$ and there can be multiple trades between a buyer and a seller to satisfy their needs. Therefore, the peer-centric configuration is represented as a bipartite graph with multiple parallel edges  in Fig.~\ref{fig:bipartite}(b), where  edges $\omega_1$, $\omega_2$, $\omega_3$ denote multiple trades of the standard size). The number of parallel edges between seller $n$ and buyer $m$ is calculated as $\min\{\overline{G}^{\mathrm p}_{n}/P,\overline{D}^{\mathrm p}_{m}/P\}$ where $\overline{G}^{\mathrm p}_{n}$ and $\overline{D}^{\mathrm p}_{m}$ are the generation capacity of seller $n$ and the maximum demand of buyer $m$. Then the number of potential matches to be considered is $\vert \Omega\vert\!=\! \sum_{n\in{\cal N}^{\mathrm s},m\in{\cal N}^{\mathrm b}}\min\{\overline{G}^{\mathrm p}_{n}/P,\overline{D}^{\mathrm p}_{m}/P\}$.

    \subsubsection{System-centric configuration}\label{subsubsec:SC}

     The system-centric configuration is modeled as follows: 
     \begin{subequations}\label{Eq:SystemCentric}
        \begin{align}
            & \max_{\Xi^{\mathrm{P2P}}} O^{\mathrm{P2P}}:=\bigg( \sum_{n \in {\cal N}^{\mathrm b}}  U_{n}(d^{\mathrm p}_n) - \sum_{n \in {\cal N}^{\mathrm s}}C_{n} (g^{\mathrm p}_n) \bigg) \label{ex_objective}\\
            & \underline{G}_{n}^{\mathrm p} \le {g}_{n}^{\mathrm p} \le \overline{G}_{n}^{\mathrm p},\quad\forall n \in {\cal N}^{\mathrm s},\label{gen_node_bound}\\
            & \underline{D}_{n}^{\mathrm p} \le {d}_{n}^{\mathrm p} \le \overline{D}_{n}^{\mathrm p},\quad\forall n \in {\cal N}^{\mathrm b},\label{demand_node_bound}\\
            & {g}_{n}^{\mathrm p} = \sum_{\omega \in \Omega_n} p_\omega,\quad\forall n \in {\cal N}^{\mathrm s},\label{gen_node_sum}\\
            & {d}_{n}^{\mathrm p} = \sum_{\omega \in \Omega_n} p_\omega,\quad\forall n \in {\cal N}^{\mathrm b},\label{demand_node_sum}\\
            & p_{\omega} \ge 0 \quad\forall \omega \in {\Omega},\label{signs} 
        \end{align}        
    \end{subequations}
    where $\Xi^{\mathrm{P2P}}=\{g^{\mathrm p}_n, d^{\mathrm p}_n, p_{\omega}, \ge 0\}$. The objective function in Eq.~\eqref{ex_objective} optimizes the welfare of all peers by maximizing the difference between the utility functions of consumers { $(U_n)$} and cost functions of producers { $(C_n)$}. Eq.~\eqref{gen_node_bound} imposes limits on the power that can be sold based on the physical limits of producer $n\in\cal{N}^{\mathrm s}$. Similarly,  Eq.~\eqref{demand_node_bound} establishes the minimum and maximum limits on the power purchased by buyer $n\in\cal{N}^{\mathrm b}$. Note that Eq.~\eqref{demand_node_bound} models elastic consumers that can adjust their consumption based on their utility function.     
    However, if the consumers are inelastic, Eq.~\eqref{demand_node_bound} can be converted into an equality by setting $\underline{D}_{n}^{\mathrm p} =\overline{D}_{n}^{\mathrm p}$. 
    Eq.~\eqref{gen_node_sum} sets the total power sold by producer $n\in\cal{N}^{\mathrm s}$ and  Eq.~\eqref{demand_node_sum} computes the total power received by consumer $n\in\cal{N}^{\mathrm b}$ from all producers, where $\Omega_n$ defines the trade set for peer $n$:
    \vspace{-1mm}
    \begin{align}\label{eq:omegan}
        & \Omega_n = 
            \begin{cases} 
                \{\omega \in \Omega \vert s(\omega) = n \},\quad \text{if}~n\in {\cal N}^{\mathrm s}, \\
                \{\omega \in \Omega \vert b(\omega) = n \},\quad \text{if}~n\in {\cal N}^{\mathrm b}. 
            \end{cases} 
    \end{align}
    Eq.~\eqref{signs} sets the power transfer from a seller to a buyer to non-negative values.  The outcome of the optimization in Eq.~\eqref{Eq:SystemCentric} yields  set of optimal matches $\Omega^*$. The system-centric optimization in Eq.~\eqref{Eq:SystemCentric} pursues the system-wide welfare-maximization at the expense of sacrificing preferences of individual peers (e.g. cost minimization for consumers or profit maximization for producers) that may act strategically in order to increase their individual welfare. Therefore, the optimization in Eq.~\eqref{Eq:SystemCentric} reminisces wholesale pool electricity markets.

    \subsubsection{Peer-centric configuration}
    Unlike the system-centric optimization in Eq.~\eqref{Eq:SystemCentric}, the peer-centric configuration matches the peers with respect to  preferences of individual peers. This process is decentralized and, therefore, can be carried out independently from the utility. As a result, each peer has the capability to negotiate, accept and reject trade $\omega$ based on their preferences, including  bounded rationality and privacy considerations \cite{BLASCH2017, Cavoukian2010}. The objective of the negotiation process is to establish a stable match between producers and consumers, i.e. there is no incentive to deviate from the cleared transactions unless the availability of producers or demand of consumers change. To obtain a stable match, we leverage the recent result by Morstyn \textit{et al.} \cite{morstyn2019bilateral} that exploits the concept of full substitutability \cite{full_substit}, which ensures that a decentralized price-adjustment process can be performed based on local information available to the peers and only requires communication between the peers engaged in trade $\omega$. Since the peer-centric configuration requires no central coordinator (e.g. the utility in the system-centric configuration), it does not necessarily achieve a welfare-maximizing solution.
    
    The stable match for each peer can be obtained as, \cite{morstyn2019bilateral}:    
    \begin{subequations}\label{Eq:Morstyn}      
    \begin{align}
        & \Omega_{n}^{*} \!=\! 
            \begin{cases}
               \! \arg\max_{\Omega_n}  \!\!\big\{ \sum_{\omega\in \Omega_n} \! \rho_{\omega}^{\mathrm s}p_\omega \!-\! C_{n}(g^{\mathrm p}_n) \big\},~\forall n \!\in\! {\cal N}^{\mathrm s},\!\!\!\!\!\!\!\\
               \! \arg\max_{\Omega_n}  \!\!\big\{ U_{n}(d^\mathrm{p}_n) \!-\! \sum_{\omega\in \Omega_n}  \rho_{\omega}^{\mathrm b}p_\omega \big\},~\forall n \!\in\! {\cal N}^{\mathrm b},\!\!\!\!\!\!\!\!
            \end{cases} \label{Selection}\\
        & \hspace{-0cm} \text{where: } \nonumber\\        
        & U_{n}(d^{\mathrm p}_n) 
        \! = \!
        \begin{cases} 
            \Upsilon_{n} (d^{\mathrm p}_n-\!\underline{D}^{\mathrm p}_{n}),\quad\text{if}~ d^{\mathrm p}_n\ge \underline{D}^{\mathrm p}_{n}\\
            -\infty,\quad \text{otherwise} 
        \end{cases}\!\!\!\!\!\!,\quad \forall n \in {\cal N}^{\mathrm b}\!,\!\label{Valuation}\\
        & {g}_{n}^{\mathrm p} = \sum_{\omega \in \Omega_n} p_\omega,\quad\forall n \in {\cal N}^{\mathrm s},\label{gen_node_sum2}\\
        & {d}_{n}^{\mathrm p} = \sum_{\omega \in \Omega_n} p_\omega,\quad\forall n \in {\cal N}^{\mathrm b}.\label{demand_node_sum2}
    \end{align}     
    \end{subequations}  
    Eq.~\eqref{Selection} is the objective function of buying/selling peer $n$ that aims to select those trades $\Omega_{n}^{*} \subseteq \Omega$ which are optimal with respect to their preferences, where $\rho_{\omega}^{\mathrm b}$ and $\rho_{\omega}^{\mathrm s}$ are the buying and selling prices of trade $\omega$. 
    Eq.~\eqref{Valuation}  is the utility function of consumer $n$ that factors in the elasticity of consumers and their willingness to adjust their consumption.
    {This utility function can be modified to reflect various preferences of peers on trades.}
    The total power generation and demand of peer $n$ are calculated  in Eq.~\eqref{gen_node_sum2} and \eqref{demand_node_sum2} as the sum of traded power $p_\omega$ over transactions $\omega \in \Omega_n$. 
    
    \textcolor{black}{Based on the policy  in Eq.~\eqref{Eq:Morstyn}, the stable match among all peers is achieved using the price adjustment process given in Algorithm 1, which is an iterative procedure that seeks consensus among all peers \cite{morstyn2019bilateral}.}         
    First, the buyer and the seller prices are initialized at zero for all trades in set $\Omega$, and saved in set of trade prices $\!\Lambda_\omega$. 
    At the beginning of each iteration, trade price set $\Lambda_\omega$ is saved as $\!\Lambda^{\mathrm{old}}_\omega$ and each peer construct their preferred trade set $\Omega^*_n$ by solving Eq.~\eqref{Eq:Morstyn}. Then for all trades that accepted by buyers but rejected by  sellers, price $\rho_\omega^{\mathrm s}$ or $\rho_\omega^{\mathrm b}$ is adjusted by value $\Delta \rho$. 
    If trade $\omega$ is selected by both seller $s(\omega)$ and buyer $b(\omega)$, then the current price is set as trade price ($\rho_\omega \!=\! \rho_\omega^{\mathrm s}\!=\! \rho_\omega^{\mathrm b}$). The adjustment process repeats until  prices for all trades converge. Once all trades are settled, the collection of such trades is returned as set $\Omega^*\!$.

    \begin{figure}[!t]
        \centering
        \includegraphics[width=\columnwidth]{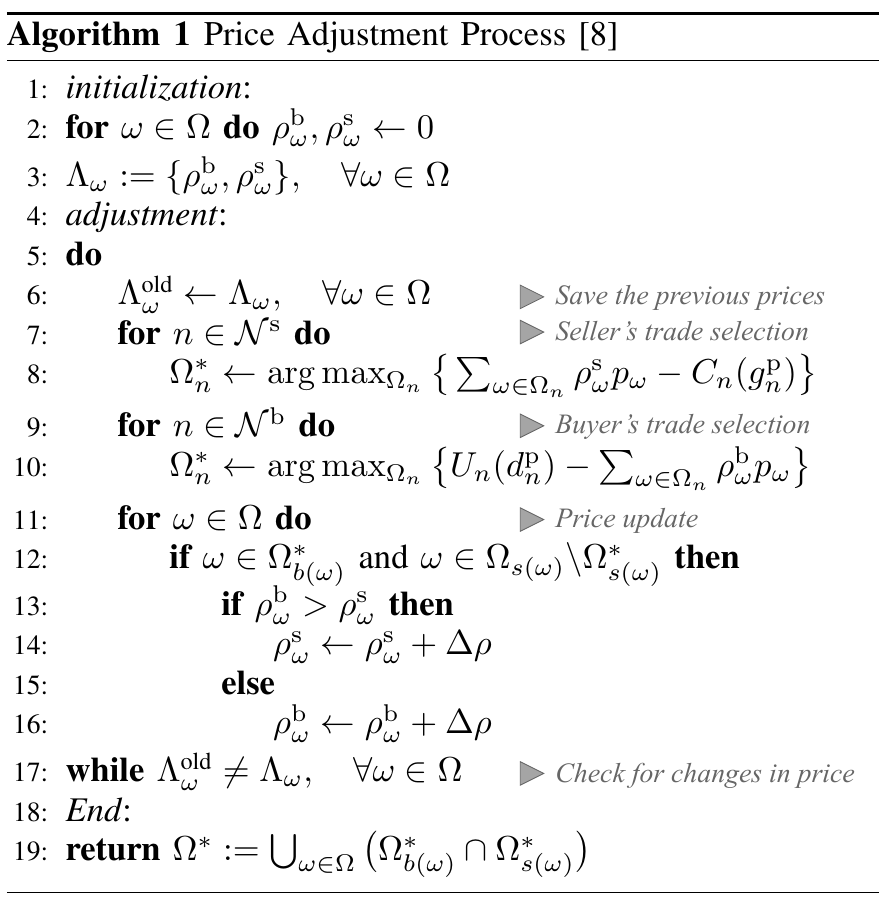}
        \label{algthm:price}        
    \end{figure}

    \subsection{Network usage charges} \label{subsec:NUC}
        Although the peer-matching configurations presented in Section~\ref{subsec:peermatching} stably generate the set of selected trades $\Omega^*$, the matching outcomes are not guaranteed to comply with distribution network limits and, therefore, it may lead to overloading the distribution system assets. To avoid this overloading, the peer-matching process must consider network constraints.  This can be accomplished by introducing network usage charges that relate the P2P interactions and the operating conditions in the distribution network. To derive these network usage charges, we use DLMPs with the intent to incentivize those P2P transactions that facilitate distribution network operations and penalize those P2P transactions that are unfavorable from the network operation perspective. 
        
     We derive the network usage charges using the second-order-cone AC OPF model,  \cite{farivar2013branch}, that scales well for large networks and makes it possible to derive DLMP components accounting for energy demand, line congestion, nodal voltage, and  power losses. Accordingly, the distribution network operations from the utility perspective can be modeled as:        
        \begin{subequations}\label{Eq:OPF}
        \begin{align}       
            & \max_{\Xi^{\mathrm{Dist}}}  O^{\mathrm{Dist}} := \sum_{b \in {\cal B}} \big[ T_b D^{\mathrm p}_b {(1-\Gamma_b)}  -C^{\mathrm u}_b p^{\mathrm g}_b \big]-C^{\mathrm w} p^{\mathrm g}_0 \label{obj_function}\\
            \begin{split}
                &(\lambda_b):\quad  f_{l \vert s(l) = b}^{\mathrm{p}} \!-\!\!\!\! \sum_{l \vert r(l) = b } \!\!\!\left( f_{l}^{\mathrm{p}} \!\!-\! a_{l} R_l \right) \!- p_{b}^{\mathrm{g}} - g_{n= b}^{\mathrm{p}} \!+\! D_{b}^{\mathrm{p}}  \!\!\!\!\!\!\!\\&\quad\quad\quad\quad~ \!+\! G_b  v_{b}=\! 0,\quad  \forall b \!\in\! {\cal B},
            \end{split}\label{P_balance}\\
            \begin{split}
                &(\mu_b):\quad  f_{l \vert s(l) = b}^{\mathrm{q}} \!-\!\!\!\! \sum_{l \vert r(l) = b } \!\!\left( f_{l}^{\mathrm{q}} \!\!-\! a_{l}  X_l \right) \!- q_{b}^{\mathrm{g}} \!+\! D_{b}^{\mathrm{q}} \!\!\!\!\!\!\!\\&\quad\quad\quad\quad\!-\! B_b v_{b} =\! 0,\quad~\forall b \!\in\! {\cal B},
            \end{split}\label{Q_balance}\\            
            & (\eta^{\mathrm +}_l):\quad \left( f_{l}^{\mathrm{p}} \right)^2 \! +\! \left( f_{l}^{\mathrm{q}}  \right)^2 \! \le \! S_l ^2 \!, \quad  \forall l \! \in \! {\cal L},\label{LineLimit_FW}\\
            & (\eta^{\mathrm -}_l):\quad \left( f_{l}^{\mathrm{p}} {}\!\! - \!\! a_{l} R_{l} \right)^2 \!\!\!+\!\! \left( f_{l}^{\mathrm{q}} \!\! - \! a_{l} X_l \right)^2 \!\! \le \!\! S_l  ^2 \! , \quad \forall l \! \in \!\! {\cal L},\label{LineLimit_BW}\\
            &  v_{o(l)}\!-\!2\!\left(R_l f_{l}^{\mathrm{p}} \!+\! X_l f_{l}^{\mathrm{q}} \right)  + a_{l} \!\left(R_l ^2 \!+\! X_l ^2 \right) \!=\! v_{r(l)},~~\forall l \! \in \! {\cal L},            
            \label{DistFlow1}\\
            & \frac{\left(f_{l}^{\mathrm{p}}\right)^2 + \left(f_{l}^{\mathrm{q}} \right) ^2 }{a_{l}} \le v_{o(l)}, \quad \forall l \! \in \! {\cal L},\label{DistFlow2}\\
            & \underline{P}_b^{\mathrm{g}} \le p_{b}^{\mathrm{g}} \le \overline{P}_b^{\mathrm{g}}, \quad \forall b \in {\cal B}, \label{PGbound}\\       
            & \underline{Q}_b^{\mathrm{g}} \le q_{b}^{\mathrm{g}} \le \overline{Q}_b^{\mathrm{g}}, \quad \forall b \in {\cal B},\label{QGbound}\\
            & \underline{V}_b \le v_{b} \le \overline{V}_b, \quad \forall b \! \in \! {\cal B}, \label{Vbound}
        \end{align}
        \end{subequations}
        where {$\Xi^{\mathrm{Dist}}=\{f^{\mathrm p}_l, f^{\mathrm q}_l, p^{\mathrm g}_b, g^{\mathrm p}_n, a_l, v_b \ge 0\}$.} Objective function $O^{\mathrm{Dist}}$ in Eq.~\eqref{obj_function} maximizes the profit of the utility given tariff $T_b$ and the amount of load served by the utility ${(1-\Gamma_b)} D_b^{\mathrm p}$ minus the cost of utility operated generators $C^{\mathrm u}_b p^{\mathrm g}_b$ and the cost of purchasing power $p^{\mathrm g}_0$  from the wholesale market at price $C^{\mathrm w}$. Note that $p^{\mathrm g}_0 = 0-f^{\mathrm p}_0$, i.e. all the power purchased in the wholesale market is injected via the root node of the distribution network. The active and reactive power balance are enforced in Eq.~\eqref{P_balance}-\eqref{Q_balance}. Unlike in the objective function in Eq.~\eqref{obj_function}, the demand enforced in  Eq.~\eqref{P_balance}-\eqref{Q_balance} accounts for  the supply from both the P2P platform and  utility, i.e. the resulting  DLMPs reflect both components. The apparent power flow limits on the receiving and sending nodes of each line are enforced in Eq.~\eqref{LineLimit_FW}-\eqref{LineLimit_BW}. Eq.~\eqref{DistFlow1} relates the line flows and nodal voltages, while Eq.~\eqref{DistFlow2} is the second-order conic constraint that convexifies the original non-convex AC OPF problem,  \cite{farivar2013branch}. 
        {This convexification is proven to be exact for distribution systems with a radial topology under rather unrestrictive assumptions (see \cite{low2014convex} for details). However, in the case of meshed distribution topologies, the second-order conic relaxation holds only under the restrictive assumption of phase shifters placed in strategic locations, \cite{low2014convex}. If this assumption does not hold, other AC power flow formulations (e.g. LinDistFlow \cite{baran1989network}) can be used  in the proposed modeling framework.}
        The active and reactive power output limits on utility generators are enforced in Eq.~\eqref{PGbound} and \eqref{QGbound}. Eq.~\eqref{Vbound}  limits nodal voltage magnitudes.

        Given the AC OPF formulation in Eq.~\eqref{Eq:OPF}, the DLMPs can be computed as follows, \cite{Papavasiliou:2017ek}:         
        \begin{align}\label{Eq:DLMP_Calc}
            & \lambda_{o(l)} =\! A_1\lambda_{r(l)} 
            \!+\!  A_2\mu_{o(l)} \!+\! A_3\mu_{r(l)} \!+\!  A_4 \eta^{\mathrm +}_{o(l)} 
            \!+\!  A_5 \eta^{\mathrm -}_{o(l)},\!\!\!
        \end{align}        
        where  parameters $A_1,\cdots,A_5$ are computed based on the optimal OPF solution, as described in Appendix. The DLMPs obtained from Eq.~\eqref{Eq:DLMP_Calc} internalizes the effects of 
        binding constraints in Eq.~\eqref{P_balance}-\eqref{LineLimit_BW} and can be interpreted in terms of 
        distribution line losses, power flow limits and nodal voltage limits, \cite{Papavasiliou:2017ek}. Given the DLMPs in Eq.~\eqref{Eq:DLMP_Calc}, we compute the network usage charge for trade $\omega$ between buyer $b(\omega)$ and seller $s(\omega)$ as follows:
        \begin{align}
            & c^{\mathrm{n}}_{\omega} = (\lambda_{b(\omega)}-\lambda_{s(\omega)})/2,\quad\forall \omega \in {\Omega}, \label{Eq:network_charge_simple}
        \end{align}
        where the factor of 2 equally splits the network usage charge between the seller and buyer. The equal allocation of the network usage charge is motivated by the assumption that the seller and buyer equally benefit from the transaction and using the distribution network. 
        \textcolor{black}{As the value of $c^{\mathrm{n}}_{\omega}$ is based on DLMPs, it reflects the network conditions and can be used to incentivize P2P trades that improve system conditions, and penalize, if otherwise.}
        Given the value of $c^{\mathrm{n}}_{\omega}$ for trade $\omega$ the buyer pays $\rho_\omega\!+\!c_\omega^{\mathrm n}$  and the seller receives the payment of $\rho_\omega - c_\omega^{\mathrm n}$.  Hence, the total network usage charges collected by the utility can be computed as: \vspace{-1mm}
        \begin{align}
            & {NUC (c_{\omega}^{\mathrm n}, p_\omega)}:=\sum_{\omega \in \Omega^*} {2c_{\omega}^{\mathrm n}}p_\omega, \label{eq:NUC}
        \end{align}  
        where $\sum_{\omega\in\Omega^*}p_\omega\!\!=\!\sum_{n\in{\cal N}^{\mathrm b}}\!D^{\mathrm p}_{n} \Gamma_{n} \!\!=\!  \sum_{n\in{\cal N}^{\mathrm s}}g^{\mathrm p}_{n}$. Note that  $NUC (c_{\omega}^{\mathrm n}, p_\omega)$  in Eq.~\eqref{eq:NUC} is  the second term in Eq.~\eqref{DSOrev_partial-decen}. 
        
        \textcolor{black}{Since the total network usage charges in Eq.~(10) are computed using DLMPs, which only recover operating costs, ${NUC (c_{\omega}^{\mathrm n}, p_\omega)}$ may generate insufficient  revenues  to support further distribution system expansion, unlike the regulated tariff discussed in Section~\ref{subsec:status-quo}. Therefore, to support this expansion, DLMPs must be extended to include capital costs, e.g. see the approach in \cite{8731686}.}

    \subsection{Coordination between the P2P platform and utility} \label{subsec:NUC_integration}
\begin{figure}[t]
            \centering
            \includegraphics[width=0.95\columnwidth]{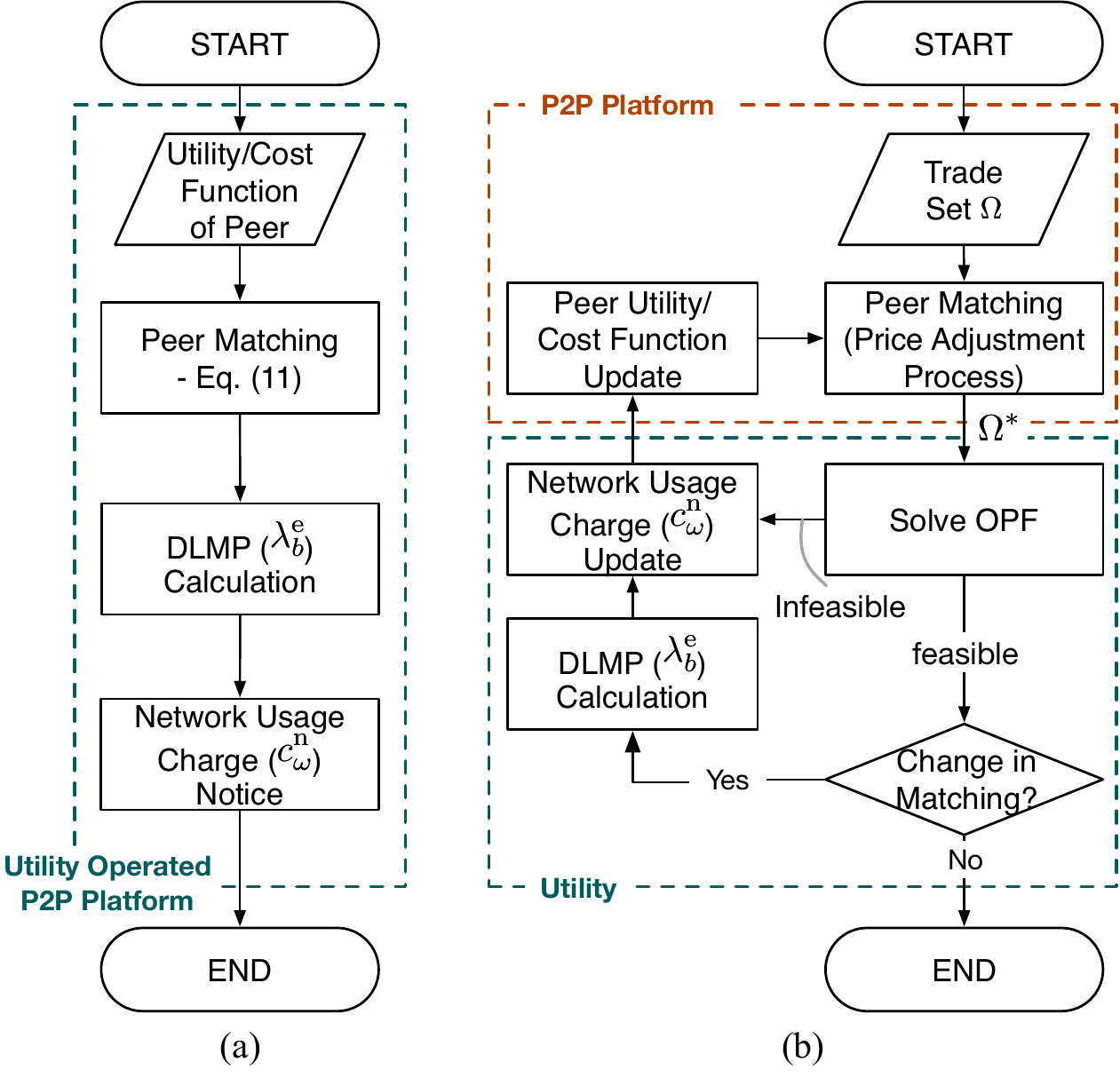}
            \caption{\small Coordination between the  proposed P2P platform and utility operations using the network usage charges for the (a) system-centric configuration and (b) peer-centric configuration. The dashed boxes delineate P2P- and utility-end procedures.}
            \label{fig:flowchart1}          
        \end{figure}

        Given the network usage charges, the P2P platform and utility operations can be coordinated to comply with distribution network limits. However, this coordination varies for the system- and peer-centric configurations. Fig.~\ref{fig:flowchart1} illustrates the coordination for each configuration as further detailed below. 
    
        \subsubsection{System-centric configuration}
        Under the system-centric P2P configuration, the integration between the P2P platform and utility operations can be achieved by co-optimizing  the P2P- and utility-end decisions. This co-optimization is shown in Fig.~\ref{fig:flowchart1}(a) and is similar to a pool market design and, therefore, imposes similar requirements on the data that peers need to share with the P2P platform (e.g. consumption and production levels, characteristics of cost and utility functions, among other preferences). Thus, the P2P platform first collects this information from peers. Second, given the collected information, the co-optimization of the P2P transaction and utility-operated distribution assets matches the peers and dispatches them to maximize the social welfare and meet distribution network constraints. This co-optimization is formulated as:
        \begin{subequations}\label{Eq:SC_final}
            \begin{align}                       
                & \max \quad O^{\mathrm{P2P}}+O^{\mathrm{Dist}} \label{SC_final_obj}\\
                & \mathrm{Eq.~} \eqref{gen_node_bound}\textendash\eqref{signs}\qquad\text{P2P constraints} \\
                & \mathrm{Eq.~} \eqref{P_balance}\textendash\eqref{Vbound}~\!\qquad\text{Network constraints}
            \end{align}
        \end{subequations}

        \noindent Third, once the co-optimization in Eq.~\eqref{Eq:SC_final} is solved, the DLMPs can be computed as in Eq.~\eqref{Eq:DLMP_Calc}. Finally, given the DLMPs, the P2P platform computes the network usage charges as in Eq.~\eqref{Eq:network_charge_simple}, and settles the transactions among the peers.   

    \subsubsection{Peer-centric configuration}
        Since the peer-centric coordination assumes that the P2P platform and the utility are operated separately, the coordination is achieved under the iterative procedure displayed in Fig.~\ref{fig:flowchart1}(b). 
        First, the P2P platform generates set of potential trades ${\Omega}$. Second, the P2P platform matches the peers, i.e. it computes set of selected trades $\Omega^*$ and the corresponding price for each trade $\rho_\omega$ using Algorithm~1. {Next, the utility solves an OPF problem in Eq.~\eqref{Eq:OPF} given the nodal injections for trades in set  $\Omega^*$ transmitted from the P2P platform.} Since the trades in set  $\Omega^*$ are myopic to network limits, solving Eq.~\eqref{Eq:OPF} may lead to a suboptimal or infeasible solution. To avoid such outcomes, the utility imposes small penalty  $\epsilon$, if  the OPF is infeasible, and the network usage charges are computed as:
        \vspace{-1mm}
        \begin{align}
            & c^{\mathrm{n}}_{\omega} = \begin{cases} (\lambda_{b(\omega)}\!\!-\!\lambda_{s(\omega)})/2,~\forall \omega \in \Omega^*, ~ \mbox{if}~\mbox{OPF}=\mbox{feasible}\!\!\!\!\!\!\!\\ \epsilon,\qquad\forall \omega \in \Omega^*, ~  \mbox{if}~\mbox{OPF}=\mbox{infeasible} \end{cases} \label{eq:network_charge}
        \end{align}
        Given the updated value of $c^{\mathrm{n}}_\omega$ and trade price $\rho_\omega$, the cost and utility functions of peers in Eq.~\eqref{Eq:Morstyn} is updated as follows: 
        \vspace{-2mm}
        \begin{align}
            & \Omega_{n}^{*} \!=\!\! 
            \begin{cases}
                \arg\max_{\Omega_n}  \big\{ \sum_{\omega\in \Omega_n}  (\rho_{\omega}^{\mathrm s}\!-\!c^{\mathrm n}_\omega)p_\omega \!-\! C_{n}(g^{\mathrm p}_n) \big\},~\forall n \in {\cal N}^{\mathrm s}\\
                \arg\max_{\Omega_n}  \big\{ U_{n}(d^\mathrm{p}_n) \!-\! \sum_{\omega\in \Omega_n}  (\rho_{\omega}^{\mathrm b}\!+\!c^{\mathrm n}_\omega)p_\omega \big\},~\forall n \in {\cal N}^{\mathrm b}
            \end{cases} \label{Selection_updated}            
        \end{align}
        With the updated cost and utility functions, the iterative procedure in Fig.~\ref{fig:flowchart1}(b) continues until the stable peer match is found.        
        Relative to the system-centric configuration, the iterative procedure allows for accommodating the negotiation process among the peers and prevents from sharing the information about individual peers with the utility.

\section{Case Study} \label{sec:casestudy}
   The case study uses the 15-bus distribution  system from \cite{Papavasiliou:2017ek} and the realistic, urban-scale  141-bus distribution feeder from \cite{khodr2008maximum}. All models are implemented using the Julia JuMP package and the code and input data are available in \cite{julia_code_gist}.

        \begin{table}[b]
            \centering
            \captionsetup{justification=centering, labelsep=period, font=footnotesize, textfont=sc}
            \caption{Peer payments and network usage charges\vspace{-0mm}}   
                \begin{center}
                \begin{tabular}{c|c|c|c|c}                
                    \hline  \hline
                    \centering  
                    \multirow{3}{*}{\!\!\!Configuration\!\!\!}&Payments &\!\!\!Revenues\!\!\!&\multirow{2}{*}{$NUC$,}&Cost of\\
                    &\!\!\!of consumers,\!\!\!&of producers,&\!\!\!\multirow{2}{*}{${\sum_\omega 2c^{\mathrm n}_\omega p_\omega}$}\!\!\!&\!\!generation,\!\!\\                    
                    &\!\!\!\!${\sum_\omega(\rho_\omega\!+\!c^{\mathrm n}_\omega) p_\omega}$\!\!\!\!&$\!\!\!\!{\sum_\omega (\rho_\omega\!-\!c^{\mathrm n}_\omega) p_\omega}\!\!\!\!$&&$\!\!\!\sum_\omega C_n(g^{\mathrm p}_n)\!\!\!\!$\\                    
                    \hline
                    \!\!\!\!\!System-centric\!\!\!\!\!&\$88.20&\$78.17&\$10.03&\$70.61\\             
                    \hline
                    \!\!\!Peer-centric\!\!\!&\$79.46&\$79.01&\$0.45&\$71.80\\
                    \hline
                    \hline              
                \end{tabular}
                \end{center}
                \label{tab:revenue15}
        \end{table}

    \begin{figure}[!b]
        \centering        
        \includegraphics[width=0.8\columnwidth]{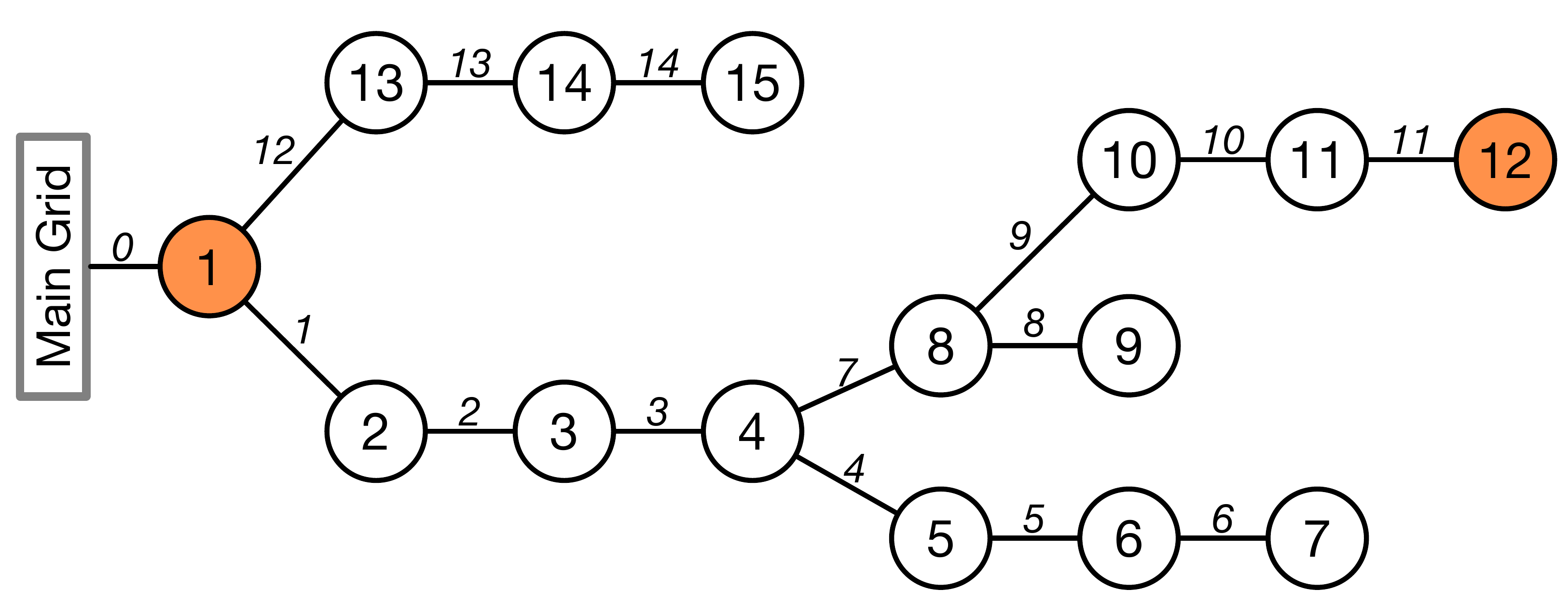}    
        \vspace{-1mm}
        \caption{\small 15-bus radial distribution system from \cite{Papavasiliou:2017ek} with nodal indices in circles. Italic numbers above edges represent line indices. Orange nodes represent producing peers.}
        \label{fig:15-bus_system}
    \end{figure}
    \subsection{15-bus distribution test system}

\begin{figure}[!t]
        \vspace{-2mm}
        \centering
        \includegraphics[width=0.91\columnwidth]{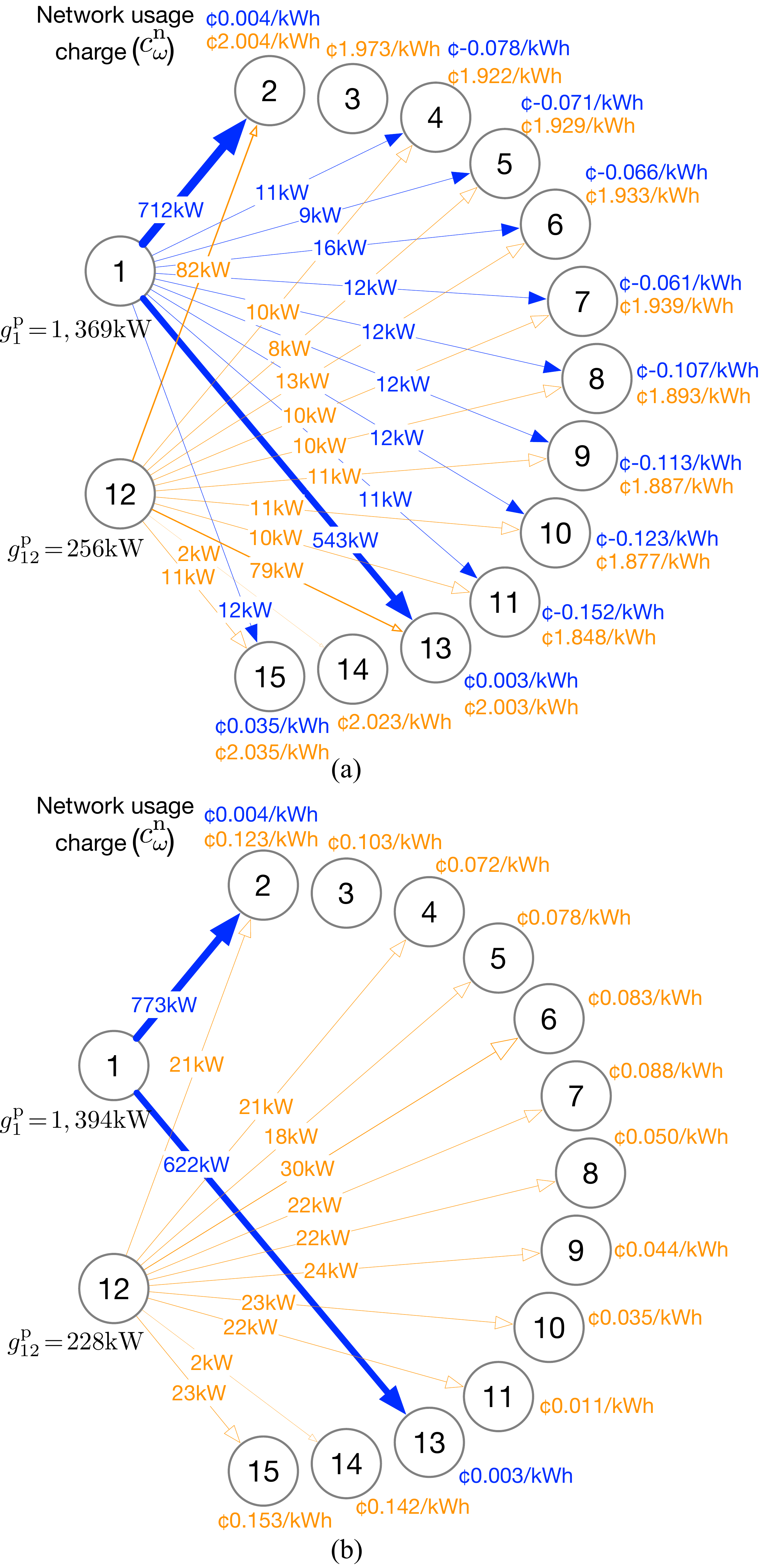}\hspace{-2mm}        
        \vspace{-3mm}
        \caption{\small P2P transactions in 15-bus distribution system under the (a)  system-centric and (b) peer-centric configurations. Blue and orange colors represent the trades by producing peers 1 and 12. The network usage charge $c^{\mathrm n}_\omega$ is given next to each consuming peers $b(\omega)$.
        {Note that there is no producer or consumer at bus 3.}}
        \vspace{-0mm}
        \label{fig:peer_15}         
        \vspace{-2mm}
    \end{figure}

        The distribution system illustrated in Fig.~\ref{fig:15-bus_system} has two producers located at node 1 and 12 with the installed capacity of 2MW and 0.4MW and the incremental cost of  \$50/MWh and \$10/MWh, respectively. The total load in the system is 1.63  MW. We select $\Gamma_b=100\%, \forall b \in \mathcal{B}$, i.e. the utility  does not supply electricity and only operates the distribution network.  Fig.~\ref{fig:peer_15} compares the P2P interactions under the system- and peer-centric configurations. While the two producing peers sell roughly the same capacity under both configurations, the sold power capacity is differently allocated among consumers. The peer-centric configuration tends to favor transactions among  neighboring buses, i.e. producer at node 1 sells exclusively to the  nodes 2 and 13 that are directly adjacent to it. On the other hand, the system-centric configuration  results in a more diverse set of the P2P interactions that allows the transactions among electrically remote nodes. 

    \begin{figure}[!t]
        \centering
        \vspace{-1mm}
        \includegraphics[width=0.92\columnwidth]{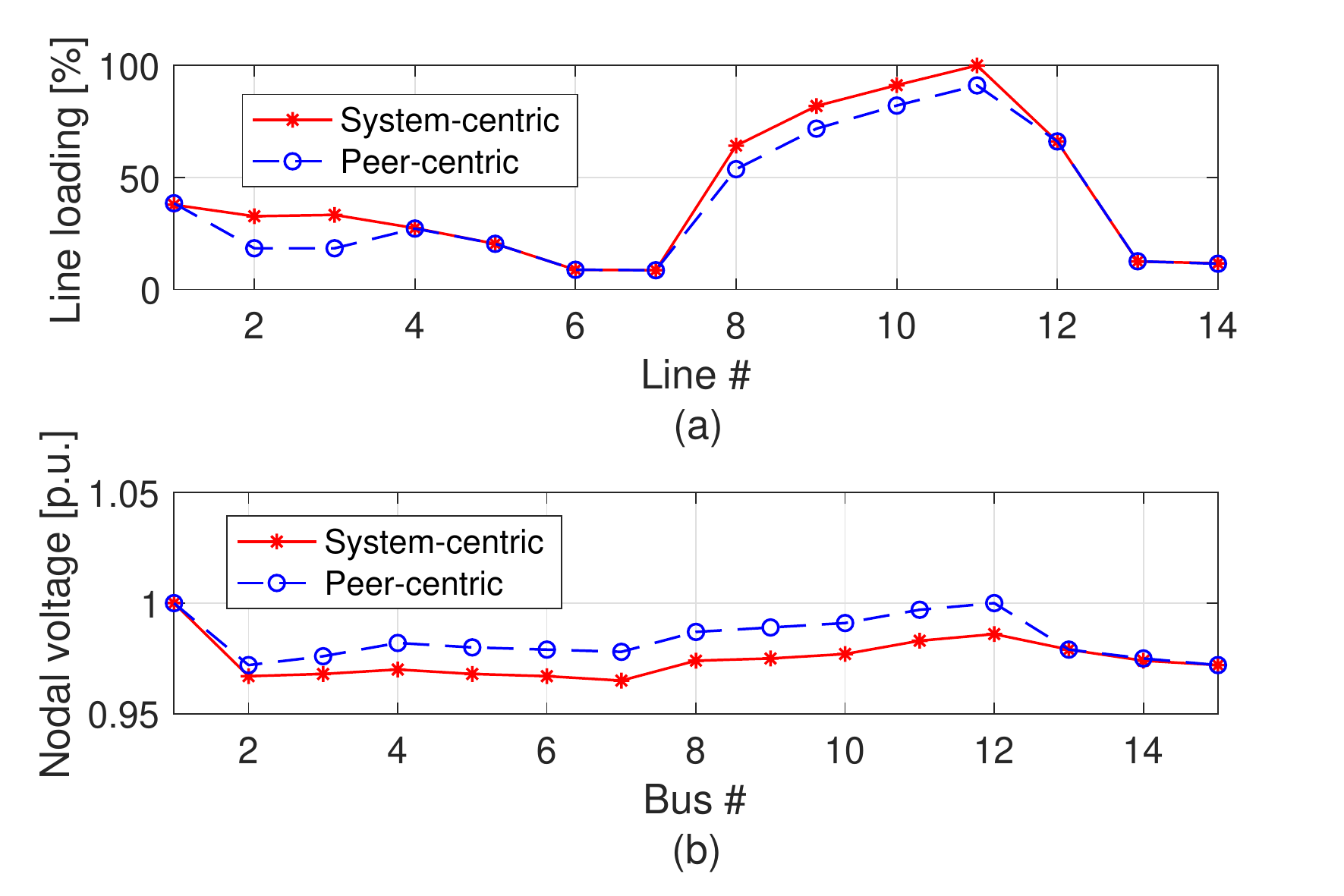}\hspace{-5mm}
        \vspace{-3mm}
        \caption{\small (a) Distribution line loading (in \% relative to $S_l$) and (b) nodal voltage magnitudes under the  system- and peer-centric configurations.}
        \label{fig:Condition_15}
        \vspace{-11mm}
    \end{figure}   

        As a result of this different trade allocation, the two P2P configurations lead to different utilization of the distribution network and, therefore, to different network usage charges. Fig.~\ref{fig:Condition_15} compares line loading and nodal voltage magnitudes under the two configurations.         
        Since the peer-centric configuration favors transactions among neighboring nodes, it does not fully utilize line capacity and incurs lower network usage charges for the peers as shown in Fig.~\ref{fig:peer_15}. 
        For example, 
        As a result, the peer-centric configuration does not maximize the social welfare of the entire distribution network.         
        On the other hand, the welfare maximum is achieved under the system-centric configuration, where the P2P transactions are explicitly co-optimized  with distribution network operations, which leads to a higher utilization of line capacity (Fig.~\ref{fig:Condition_15}) and greater network usage charges in total (Fig.~\ref{fig:peer_15}). 
        {
        Since network usage charges represent the system conditions resulting from the settled trades, they can attain both positive and negative values. 
        For example, some trades in Fig.~\ref{fig:peer_15} have negative network usage charges, which implies that such trades would relieve line congestion or improve a nodal voltage profile. Purchasing electricity from peer 1 with the incremental cost of \$50/MWh would be more expensive than peer 12 with the incremental cost of \$10/MWh, but such trades could improve network conditions, and therefore, network usage charges are discounted.}            
        These differences in utilization of the distribution network lead to different network usage charges and achieve different levels of social welfare as compared in Table~\ref{tab:revenue15}. Accordingly, the system-centric configuration results in a greater revenue of the utility from the network usage charges, while the peer-centric formulation ensures the least-cost electricity supply. 

    \subsection{141-bus urban-scale distribution feeder}

    \begin{table}[!b]
        \centering        
        \captionsetup{justification=centering, labelsep=period, font=footnotesize, textfont=sc}
        \caption{location, incremental cost and capacity of DGs\vspace{-3mm}}   
        \begin{center}
            \begin{tabular}{c|c|c|c|c|c|c|c|c|c}                
                \hline  \hline
                \centering                  
                Peer    &1   &30  &40  &50  &60  &70  &80  &101 &121\\
                \hline
                Incremental &\multirow{2}{*}{20}  &\multirow{2}{*}{10}  &\multirow{2}{*}{12}  &\multirow{2}{*}{11}  &\multirow{2}{*}{15}  &\multirow{2}{*}{10}  &\multirow{2}{*}{17}  &\multirow{2}{*}{10}  &\multirow{2}{*}{13}\\
                Cost, \$/MWh &&&&&&&&&\\
                \hline
                Capacity, MW    &2   &2   &2   &2   &2   &2   &2   &2   &2\\                
                \hline
                \hline              
            \end{tabular}
        \end{center}
        \label{tab:DGinfo}
        \vspace{-0mm}
    \end{table}

    We add  9 DERs with parameters given in Table~\ref{tab:DGinfo} to the 141-bus distribution system from \cite{khodr2008maximum}, which has  the total load of $11.98\text{MW}$. Different penetration levels of P2P transactions are simulated by varying parameter $\Gamma_b$ between 0 and 0.6, while assuming that the utility must operate the distribution network and supply the residual demand.
   \begin{figure}[!b]
        \centering
        \hspace{0.3mm}
        \includegraphics[width=0.98\columnwidth]{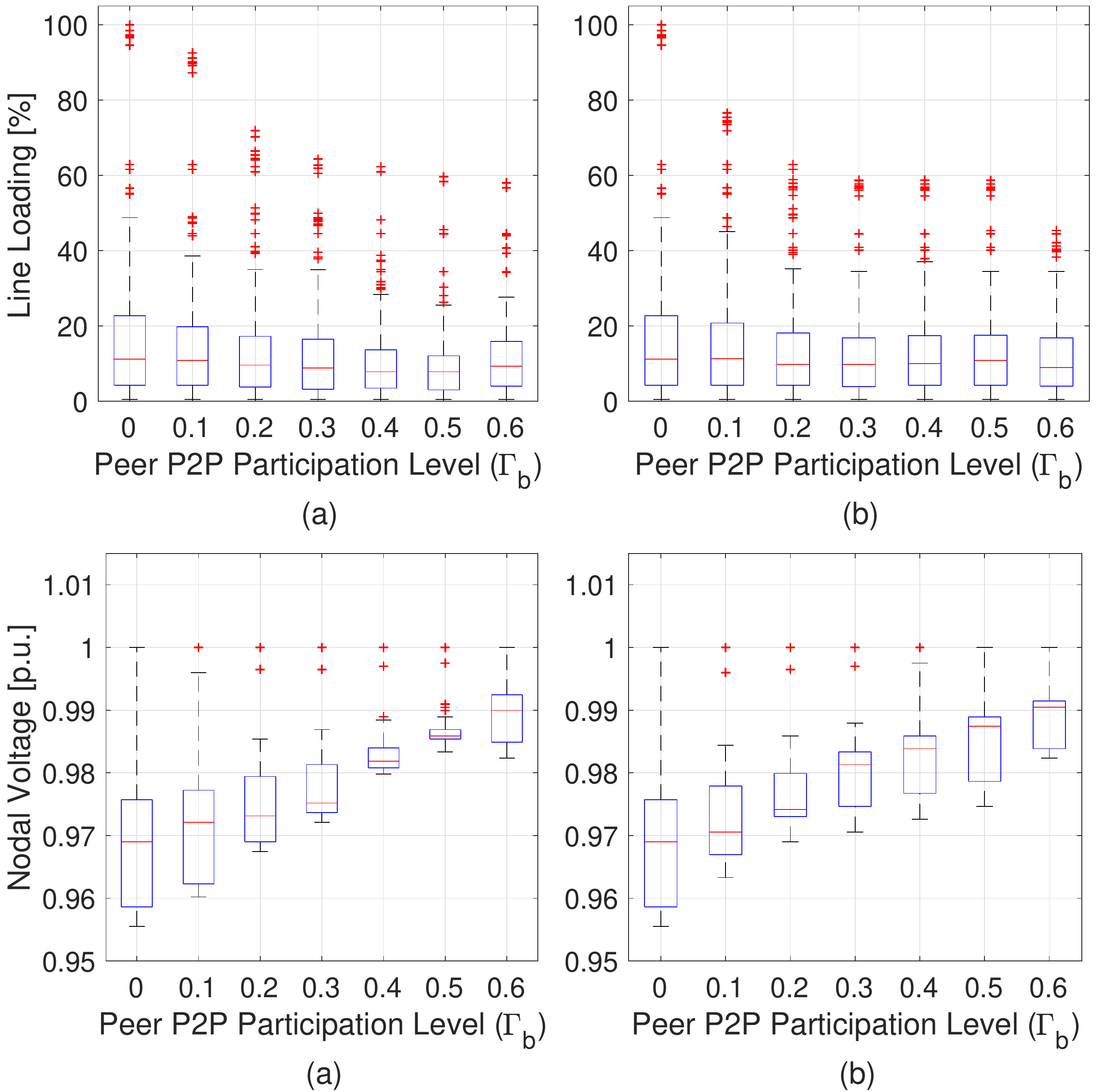}
        \vspace{-3mm}
        \caption{\small Comparison of line loading  (in \% relative to $S_l$) and voltage magnitudes under the (a) system-centric and (b) peer-centric configurations.  The red line within the blue box represents the median value, the bottom and top edges of the box represent the first and third quartiles and the outliers are plotted outside the box in red.}
        \label{fig:Vboth_141}          
    \end{figure}    
    \begin{figure}[t]
        \vspace{-4mm}
        \centering
        \hspace{2mm}\includegraphics[width=0.98\columnwidth]{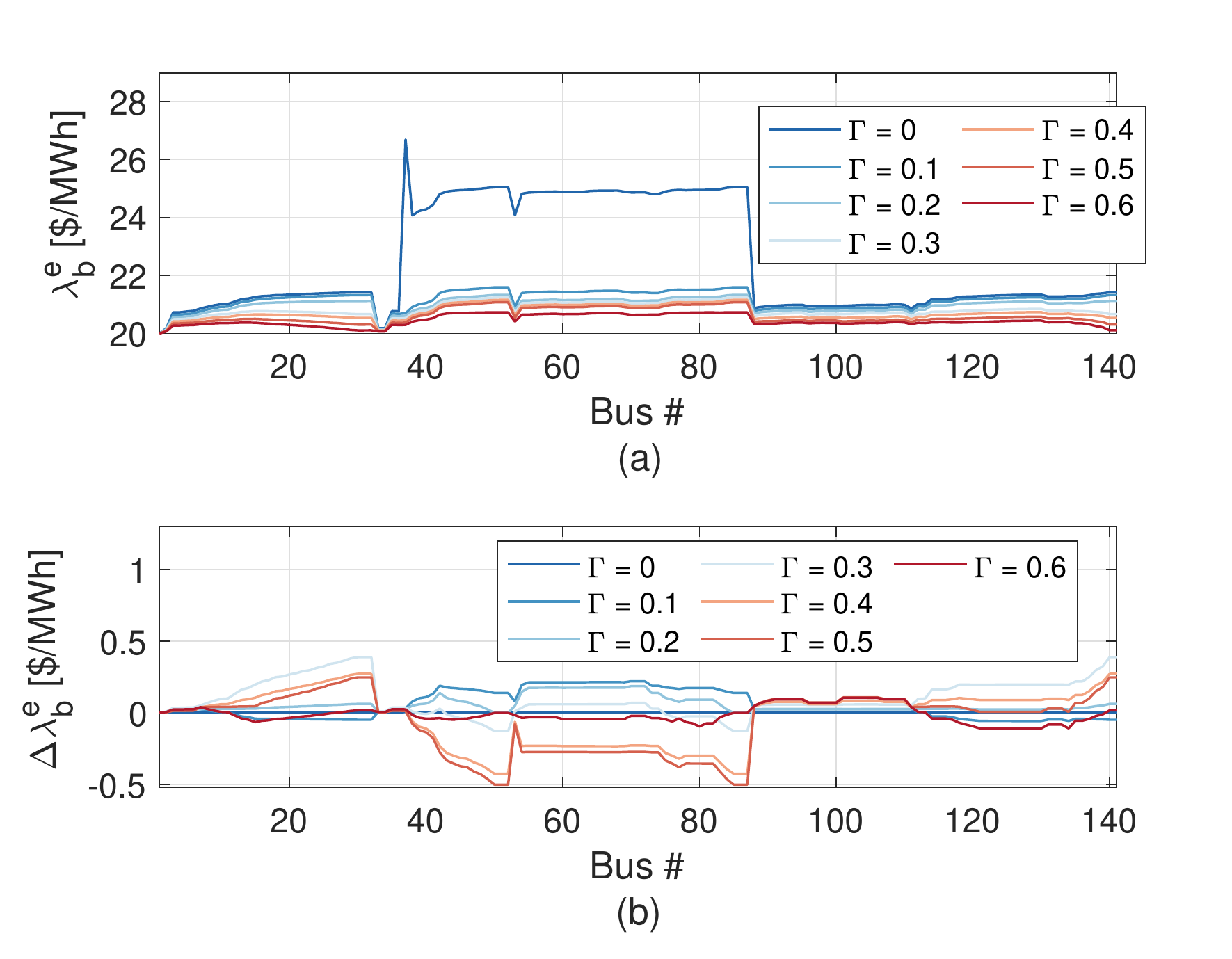}\hspace{-2mm}
        \vspace{-4mm}        
        \caption{\small (a) DLMPs of the system-centric configuration and (b) the difference of DLMPs between the system- and peer-centric configurations ($\Delta\lambda_b\!=\!\lambda^{\mathrm{SC}}_b\!-\!\lambda^{\mathrm{PC}}_b$) with  $\Gamma_b\!=\!\Gamma$.
        }
        \label{fig:DLMP_both141}
    \end{figure}
    Fig.~\ref{fig:Vboth_141} compares the line loading and voltage magnitudes under the two P2P configurations. As the P2P penetration level increases in both cases, the line loading and its variance across lines monotonically reduce. This reduction is mainly achieved due to the fact that the P2P interactions offset centralized electricity production by local power injections. Similarly, local power injections improve a voltage profile across the distribution network due to the reduction of power losses.  As a result, the P2P interactions under both configurations lead to sizable reductions in the magnitude and volatility of DLMPs as the P2P penetration level increases, as shown in Fig.~\ref{fig:DLMP_both141}. The difference in  DLMPs under the system- and peer-centric configurations merely exists and  further reduces  as the P2P platform penetration level increases.  
   
    Table~\ref{tab:revenue} compares average network usage charge ${\mathbb{E}}(c^{\mathrm n}_\omega)\!=\!{\sum_{\omega\in\Omega^*}c^{\mathrm n}_\omega p_\omega}/{\sum_{\omega\in\Omega^*} p_\omega}$ of the system- and peer-centric configurations with different P2P platform participation levels $\Gamma_b$ and corresponding P2P supplied demand $\sum_{b\in{\cal B}}\Gamma_b D^{\mathrm p}_b$. The system-centric yields higher network usage charges, because it tends to spread the use of P2P resources across the entire distribution network, which maximizes the global welfare.  On the other hand, the peer-centric configuration results in lower and, even negative in some cases, network usage charges, because it favors network-friendly P2P trades without accounting for the welfare-maximization.

    \begin{table}[t]
    \centering
    \vspace{-0mm}
    \captionsetup{justification=centering, labelsep=period, font=footnotesize, textfont=sc}
    \caption{Average network usage charges\vspace{-4mm}}   
        \begin{center}
        \begin{tabular}{c|c|c|c|c|c|c|c|c}                
            \hline  \hline                                
            \!\!P2P supplied\!\!&$\Gamma_b\!$&\!0\!&0.1&0.2&0.3&0.4&0.5&0.6\!\!\!\\
            \cline{2-9}
            Demand&$\!\!\!\sum_b\!\Gamma_b\! D^{\mathrm p}_b\!\!\!\!$&\!0\!&1.20&2.40&3.59&4.79&5.99&7.19\!\!\!\\
            \hline                
            \multirow{2}{*}{$\!\!\!\!{\mathbb{E}}(c^{\mathrm n}_\omega),\text{\$/MWh}$\!\!\!\!}&\!\!\!System\!\!\!&\!0\!&\!0.65\!    &\!0.84\!    &\!0.84\!    &\!0.93\!    &\!0.94\!    &\!0.93\!\!\!\\
            \cline{2-9}                
            &\!\!Peer\!\!&\!0\!&\!\!-0.09\!\!&\!\!-0.07\!\!&\!\!-0.03 \!\! &\!\!-0.03 \!\!&\!\!-0.02\! \!&\!\!0.03\!\!\!\!\\
            \hline
            \hline              
        \end{tabular}
        \end{center}
        \label{tab:revenue}                
        \vspace{-5mm}
    \end{table}

\section{Conclusion} \label{sec:conclusion}    
    This paper describes a new distribution grid architecture with a P2P platform enabling trades among small-scale DERs. The proposed P2P platform can internalize both of the system- and peer-centric matching processes. The system-centric configuration achieves a centralized peer matching in a welfare-maximizing manner and the peer-centric configuration allows peers to reflect their preferences and match in a decentralized way. For both configurations, we use  DLMPs to coordinate the distribution system operation and the P2P energy trading and to design network usage charges that peers pay for using the distribution network operated by the utility. Our simulations analyze  techno-economic performance of the P2P platform  from the perspective of the utility and peers. 
    
    \textcolor{black}{The results presented in this paper point to multiple directions for our future work. First, it is important to extend the proposed P2P architecture to account for dynamically changing utility and cost functions of peers, as well as their ability to observe and collect additional information about the distribution system and other peers. This information, in turn, can be used by peers strategically to advance their self-interest at the expense of other P2P participants, which should be mitigated. Second, it is important to internalize various demand- and supply-side uncertainties to include their effects on peer matching and network usage charges. Third, the AC power flow model can be extended to accommodate meshed distribution system typologies. Furthermore, the power flow model can be extended to compute network usage charges in the presence of three-phase, unbalanced  operating conditions, which often occur in low-voltage distribution systems. Finally, since the proposed P2P architecture changes operating principles, long-term planning methods should be modified accordingly. In particular, these methods must account for  decentralized decision-making processes executed by peers and internalize the capital cost of system expansion in  network usage charges.}
    
    \bibliographystyle{IEEEtran}
    \bibliography{P2P}

    \enlargethispage{-30mm}
    
\appendix
\section{Appendix} \label{subsec:appendixDLMP}

As derived in \cite{Papavasiliou:2017ek}, the DLMPs are computed for the origin node $o(l)$ of line $l$ as follows:
    \begin{subequations}
        \begin{align}\label{Eq:DLMP_Calc_Detail}
            & \hspace{-3.5mm}\lambda_{o(l)} =\! A_1\lambda_{r(l)} 
            \!+\!  A_2\mu_{o(l)} \!+\! A_3\mu_{r(l)} \!+\!  A_4 \eta^{\mathrm +}_{o(l)} 
            \!+\!  A_5 \eta^{\mathrm -}_{o(l)}\!\!,
        \end{align}
        where  $\mu_{o(l)}$,  $\mu_{r(l)}$,  $\eta^{\mathrm +}_{o(l)}$  and $\eta^{\mathrm -}_{o(l)}$  are the dual variables of the optimization in Eq.~\eqref{Eq:OPF} and parameters $A_1$, $A_2$, $A_3$, $A_4$, $A_5$ denote the following functional expressions:
        \begin{align}
            & \hspace{-3.5mm}   A_1 
                \!=\! \frac{((f^{\mathrm p}_l)^2\!+\!(f^{\mathrm q}_l)^2)X_l\!+\!a_l f^{\mathrm q}_l (R_l^2\!-\!X_l^2)\!-\!2a_l f^{\mathrm p}_l R_l X_l}{((f^{\mathrm p}_l)^2+(f^{\mathrm q}_l)^2){X_l} - a_l f^{\mathrm q}_l (R_l^2+X_l^2)}
            \\      
            & \hspace{-3.5mm} A_2 
                =   \frac{((f^{\mathrm p}_l)^2+(f^{\mathrm q}_l)^2){R_l}-a_lf^{\mathrm p}_l(R_l^2+X_l^2)}{((f^{\mathrm p}_l)^2+(f^{\mathrm q}_l)^2){X_l}-a_lf^{\mathrm q}_l(R_l^2+X_l^2)}
            \\
            & \hspace{-3.5mm} A_3 
                \!= \!\frac{-((f^{\mathrm p}_l)^2\!+\!(f^{\mathrm q}_l)^2){R_l}\!+\!a_lf^{\mathrm p}_l\!(R_l^2\!-\!X_l^2)\!+\!2a_lf^{\mathrm q}_lR_l X_l}
                {((f^{\mathrm p}_l)^2\!+(f^{\mathrm q}_l)^2){X_l}-a_lf^{\mathrm q}_l(R_l^2+X_l^2)}\\
            & \hspace{-3.5mm} A_4 
                 = \frac{2((f^{\mathrm q}_l)^3{R_l}-(f^{\mathrm p}_l)^3{X_l})+2f^{\mathrm p}_lf^{\mathrm q}_l(f^{\mathrm p}_l{R_l}-f^{\mathrm q}_l{X_l})}{((f^{\mathrm p}_l)^2+(f^{\mathrm q}_l)^2){X_l}-a_lf^{\mathrm q}_l(R_l^2+X_l^2)}\\
            \begin{split}
                & \hspace{-3.5mm} A_5
                = \frac{2((f^{\mathrm q}_l)^3{R_l}-2(f^{\mathrm p}_l)^3{X_l})+2f^{\mathrm p}_lf^{\mathrm q}_l(f^{\mathrm p}_l{R_l}-f^{\mathrm q}_l{X_l})}{((f^{\mathrm p}_l)^2+(f^{\mathrm q}_l)^2){X_l}-a_lf^{\mathrm q}_l(R_l^2+X_l^2)} \\
                & \quad\quad+ \frac{2a_l^2(f^{\mathrm q}_lR_l^3-f^{\mathrm p}_lX_l^3)-4a_lf^{\mathrm p}_lf^{\mathrm q}_l(R_l^2-X_l^2)}{((f^{\mathrm p}_l)^2+(f^{\mathrm q}_l)^2){X_l}-a_lf^{\mathrm q}_l(R_l^2+X_l^2)}\\
                & \quad\quad+ \frac{4a_lR_l X_l((f^{\mathrm p}_l)^2-(f^{\mathrm q}_l)^2)}{((f^{\mathrm p}_l)^2+(f^{\mathrm q}_l)^2){X_l}-a_lf^{\mathrm q}_l(R_l^2+X_l^2)}\\
                & \quad\quad+ \frac{-2a_l^2R_l X_l(f^{\mathrm p}_l{R_l}-f^{\mathrm q}_l{X_l})}{((f^{\mathrm p}_l)^2+(f^{\mathrm q}_l)^2){X_l}-a_lf^{\mathrm q}_l(R_l^2+X_l^2)}\!\!\!
            \end{split}
        \end{align}
    \end{subequations}

    % \vfill
    \begin{IEEEbiographynophoto}{Jip Kim} received his B.S. degree in Electrical and Electronic Engineering from Yonsei University, Seoul, Korea, in 2012 and M.S. degree in Electrical and Computer Engineering from Seoul National University, Seoul, Korea, in 2014. He is currently a Ph.D. student in the Smart Energy Research (SEARCH) Group and an Ernst Weber Fellow at the Department of Electrical and Computer Engineering, New York University, NY, USA. His research focuses on developing mathematical models and optimization algorithms to solve power system engineering problems. 
    \end{IEEEbiographynophoto}
    % \vfill
    \begin{IEEEbiography}{Yury Dvorkin} received the Ph.D. degree from the University of Washington, Seattle, WA, USA, in 2016. He is currently an Assistant Professor  and Goddard Faculty Fellow in the Department of Electrical and Computer Engineering, New York University, NY, USA, with a joint appointment at the New York University's Center for Urban Science and Progress. Dvorkin was  awarded the Scientific Achievement Award by Clean Energy Institute (University of Washington) for his doctoral dissertation in 2016,  the 2019 NSF CAREER Award and  2019 Goddard Junior Faculty Award (New York University).  His research interests include short- and long-term planning in power systems with renewable generation and power system economics. Dvorkin is an Associate Editor of the IEEE Transactions on Smart Grid.
 
    \end{IEEEbiography}
    \vfill

\end{document}